# Two-dimensional manganese oxide nanolayers on Pd(100): Surface phase diagram

F. Li[1], G. Parteder[1], F. Allegretti[1], C. Franchini[2], R. Podloucky[3] , S. Surnev[1], F.P. Netzer[1]

[1] Institute of Physics, Surface and Interface Physics, Karl-Franzens University Graz, A-8010 Graz, Austria

[2] Faculty of Physics, University of Vienna and Center for Computational Materials Science, A-1090 Vienna, Austria

[3] Institute of Physical Chemistry, University of Vienna and Center for Computational Materials Science, A-1090 Vienna, Austria

E-mail: francesco.allegretti@uni-graz.at
falko.netzer@uni-graz.at

**Abstract:**

Two-dimensional manganese oxide layers have been grown on Pd(100) and have been characterized by scanning tunnelling microscopy (STM), low-energy electron diffraction (LEED) and X-ray photoelectron spectroscopy (XPS). The complex surface phase diagram of $MnO_x$ on Pd(100) is reported, where nine different novel Mn oxide phases have been detected as a function of the chemical potential of oxygen $\mu_O$. Three regions of the chemical potential of oxygen can be identified, in which structurally related oxide phases are formed, often in coexistence at the surface. The different regions of $\mu_O$ are reflected in the oxidation states of the respective Mn oxide nanolayers as revealed by the Mn 2p and O 1s XPS binding energies. The $MnO_x$ nanolayers form two-dimensional wetting layers and it is speculated that they mediate the epitaxial growth of MnO on Pd(100) by providing structurally graded interfaces.

## 1. Introduction

Manganese oxides are distinguished by a rich variety of structural, electronic and magnetic properties that have rendered them materials of interest for numerous applications. Based on their particular physical and chemical properties, areas of applications include heterogeneous catalysis, electrochemistry, environmental waste treatment, or novel electronic device technology [1-4]. As the parent compounds of the manganites, the class of complex oxides which are famous for their giant magnetoresistance effect [5], they also display a variety of interesting magnetic and electronic properties [6]. As for other transition metal oxide materials, thin films of manganese oxides supported on metal substrates bear promise for technological use and fundamental scientific studies. Thin films of MnO have been prepared with epitaxial order by various deposition methods on different noble metal single crystal surfaces, such as Ag(001) [7], Rh(100) [8], and Pt(111) [9, 10] surfaces. Recently, we have grown epitaxial MnO films on Pd(100) with either (100) or (111) surface orientation [11], the orientation depending on the specific kinetic parameters during growth. The epitaxial growth of MnO on all these noble metal surfaces is surprising in view of the large lattice mismatch that exists between the bulk rocksalt crystal structure of the MnO overlayer and the metallic substrates. Indeed, the lattice mismatch between Ag(001), Pt(111), and Pd(100) and MnO is ~9%, ~13% and ~14%, respectively. In the present work we have set out to investigate this question of epitaxial growth in the presence of such a large lattice mismatch. We find that ultrathin layers of variable $MnO_x$ stoichiometry, only 1-2 monolayers thick, can be formed at the metal-MnO interface and argue that they may mediate the epitaxial growth by providing a structurally graded interface. In order to characterize these interfacial $MnO_x$ nanolayers structurally and chemically, we have transferred them from the interface to the surface and have studied the formation of two-dimensional (2D) Mn oxide monolayer structures at the surface of Pd(100) under various kinetic growth conditions.

Here, we address the multitude of different Mn oxide nanolayer phases found on Pd(100) as a function of the chemical potential of oxygen and characterize their structural properties by scanning tunnelling microscopy (STM) and low-energy electron diffraction (LEED). The atomistic models of the different phases as derived by combining the experimental high-resolution STM and vibrational electron energy loss spectroscopy (HREELS) data with ab initio density functional theory calculations will be presented in a forthcoming publication [12]. We find at least nine different 2D $MnO_x$ phases on Pd(100), which are novel in terms of the known Mn oxide bulk crystal structures and which are stabilized by the metal-oxide interface and the confinement in the direction perpendicular to the surface. In this paper, we introduce the manifold Mn oxide nanophases and describe an experimental phase diagram in terms of phase stability regions versus the chemical potential of oxygen, as established by STM, LEED and X-ray photoelectron spectroscopy.

**2. Experimental**

STM and LEED experiments have been performed in a custom-designed ultrahigh vacuum (UHV) chamber operated at a base pressure below $2\times10^{-10}$ mbar, equipped with a variable-temperature atomic force-scanning tunnelling microscope (Omicron VT-AFM/STM, Germany), a four-grid LEED optics, an electron beam evaporator, a quartz crystal microbalance, and sample heating and cleaning facilities. The sample can be transferred with a magnetic transfer rod between the different preparation and analysis stages in the UHV chamber. The STM images were recorded in a constant current mode at room temperature. Electrochemically etched W tips were used, which have been cleaned *in situ* by electron bombardment.

The LEED measurements have been complemented by spot-profile analysis low energy electron diffraction (SPA-LEED) experiments performed in another UHV chamber, described in detail elsewhere [13]. High-resolution X-ray photoemission measurements (HR-XPS) with use of synchrotron radiation were carried out at beamline I311 in the Swedish synchrotron radiation laboratory MAX-lab in Lund [14]. Photon energies of 620 eV and 800 eV have been used for exciting electrons from the O 1s and Mn 2p core levels, respectively, and the corresponding experimental resolution was better than 250 meV at hν = 620 eV and 300 meV at hν = 800 eV. The core-level spectra were measured at room temperature and at normal emission. The binding energy scale was calibrated with respect to the Fermi energy of the crystal in each case. The core-level spectra were normalized to the secondary electron background at a few eV lower binding energy than the respective core level peak.

Clean Pd(100) surfaces were prepared by 1.5 keV $Ar^+$-ion sputtering, followed by annealing to 1000 K for several minutes, and by heating cycles in $O_2$ atmosphere ($2\times10^{-7}$ mbar) at 570 K followed by a final short flash to 1000 K in UHV. Manganese oxide layers have been prepared by either reactive evaporation (RE) of Mn metal onto the clean Pd(100) surface in an oxygen atmosphere or by post-oxidation (PO) of Mn metal films. The PO procedure was found to result in more atomically smooth layers, consisting of a predominantly single oxide phase and has been therefore preferred over the RE one. Here, Mn layers were first deposited in UHV at room temperature (300 K) on the Pd(100) surface and subsequently oxidized in an oxygen atmosphere, where the oxygen pressure was varied between $5\times10^{-8}$ mbar and $5\times10^{-6}$ mbar and the sample temperature was between 600 K and 800 K. The Mn deposition rate was monitored by a quartz crystal microbalance and typically an evaporation rate of 0.2 monolayer/min was employed. The Mn oxide coverage is given in monolayers (ML), where 1 ML contains $1.32\times10^{15}$ Mn atoms/cm$^2$, which is equal to the Pd(100) surface density. In the present work we restrict ourselves to Mn-oxide layers in the coverage region up to 1 ML.

In order to reproduce the combined effect of the temperature $T$ and the oxygen pressure $p$ on the Mn-oxide surface phase diagram we plot the latter (see section 4) as a function of the chemical potential of oxygen $\mu_O(T, p)$, using the expression derived by Reuter and Scheffler [15]:

$$\mu_O(T,p) = \mu_O(T,p^o) + 1/2 kT \ln\left(\frac{p}{p^o}\right),$$

with the values of $\mu_O(T,p^o)$ tabulated in [15] at $p^o$=1 atm.

## 3. Results and discussion

The various oxide phases discussed in the following are all characterized by specific windows in the parameter space of the thermodynamic variables temperature and oxygen pressure. Some of the phases are stable only in a very narrow range of the parameters and this often causes the coexistence of phases that are adjacent in the phase diagram. To introduce some systematics in the presentation we will keep the oxide coverage in the range 0.75-1 ML. The sample temperature during the oxidation process has been varied in limited ranges of values: 600-700 K and 700-800 K at high and low oxygen pressure conditions, respectively. The variation of the chemical potential of oxygen during oxide formation is then mainly effectuated by varying the pressure of oxygen during evaporation or post-oxidation; the oxygen pressure can be converted into the chemical potential of oxygen using the standard thermodynamic expressions (see above).

*3.1 The hexagonal $MnO_x$ phases at $5 \times 10^{-6}$ mbar $> p_{O2} > 5 \times 10^{-7}$ mbar*

In this, what we call here the "oxygen-rich", pressure regime three Mn oxide phases have been detected, which are characterized by a hexagonal or quasi-hexagonal symmetry in STM and LEED. Figure 1(a) shows a wide scan STM image of the Pd(100) surface after evaporation of Mn with an oxygen pressure in the range $5 \times 10^{-6} - 1 \times 10^{-6}$ mbar (670 K substrate temperature). Two domains consisting of stripe structures are recognized, which are rotated by ~90° with respect to each other. The higher resolution STM image 1(c) reveals a complicated pattern of stripes of maxima, which are separated periodically by darker troughs. The periodicity of the troughs can be measured from the line scan across the troughs (figure 1(d)) to 22-23 Å. In the high-resolution STM image of the insert of figure 1(c) a quasi-hexagonal arrangement of protrusions is recognized with a lattice constant of ~3.1 ± 0.1 Å. The LEED pattern from this surface (figure 1(b)) displays reflections close to the Pd(100) (10) and (01) substrate positions, which are elongated along the direction of the corresponding reciprocal lattice vectors, and other reflections in between the (01), (10) and (11) substrate positions that are elongated in the direction perpendicular to it: this is indicative of a quasi-hexagonal overlayer lattice. Figures 2(a) and (b) show a real space model of the overlayer (dots) on the Pd(100) lattice (grid of lines) and the corresponding simulated reciprocal space diffraction pattern [16]. Clearly, this overlayer model reflects the correct LEED pattern as experimentally observed in figure 1(b). According to this model, the overlayer lattice vectors are $\mathbf{b_1}$ = 2.94 Å and $\mathbf{b_2}$ = 3.14 Å, and this Mn oxide phase will be designated as HEX-I *distorted*. The overlayer is incommensurate in the $\mathbf{b_1}$ direction, but in the $\mathbf{b_2}$ direction a row matching condition exists (see figure 2(a)). The STM image of figure 1(c) may then be understood in terms of a Moiré pattern as a result of the interference of the Mn oxide overlayer lattice with the square Pd(100) surface lattice ($a_1 = a_2 = a_{Pd}$ = 2.75 Å). Taking a ratio of $b_1/a_{Pd} \approx 16/15$, a Moiré pattern can be constructed in a geometrical model that displays modulations in form of broad lines inclined with respect to the Pd [110] direction ($a_1$) and with an average periodicity of ~22 Å parallel to Pd [110] (see figure 2(c)). In this model, only one type of overlayer atoms is considered for simplicity, and a different colour gradation (white, grey, black) is used to highlight the different lateral registry of these overlayer atoms relative to the underlying matrix of substrate atoms. For on-top/bridge location, the white/black colour of the overlayer atoms reflects the different height above the surface. Along the $a_1$ direction the same site registry is obtained for every $16 \times a_{Pd}$ lattice constants, but when including the registry along lines inclined by 60° with respect to the $a_1$ direction a site registry periodicity of $8 \times a_{Pd} = 22$ Å on average is observed. The resulting modulation (figure 2(c)) is remarkably similar to the experimental STM image as seen in figure 1(c). Implicitly, this suggests an oxide film consisting of alternately stacked layers with quasi-hexagonal symmetry containing only one single atom species, as it is realized in the MnO(111) structure. This will be discussed in detail in reference [12].

Under similar preparation conditions as discussed above, another Mn oxide phase has been occasionally observed, which appears to be closely related to the *distorted* HEX-I. A LEED picture of this phase is reproduced in figure 3(a). Here, a perfectly hexagonal pattern is recognized, which corresponds to a lattice with $b_1 = b_2 = 3.14$ Å, and which will be referred to as *undistorted* HEX-I structure. Unfortunately, no good quality STM images have been obtained from this phase. Figures 3(b) and (c) show the real space model and the simulated diffraction pattern of the *undistorted* HEX-I [16], respectively; two orthogonal domains contribute to the pattern. The fact that the *undistorted* HEX-I is less frequently observed than the *distorted* HEX-I seems to indicate that the former structure is less stable and requires particular conditions for kinetic stabilization. We speculate that the larger lattice mismatch (14 % against 7%) in the $b_1$ direction and the related interfacial strain is responsible for the lower stability of the *undistorted* HEX-I as compared to the *distorted* one.

At somewhat lower oxygen pressure, in the range $1 \times 10^{-6} > p_{O2} > 5 \times 10^{-7}$ mbar, a third hexagonal Mn oxide phase (HEX-II) can be observed. Figures 4(a) and (b) display STM images of this HEX-II structure, which is characterized by a hexagonal array of triangular shaped maxima and thus has a very different appearance in the STM as the HEX-I phase. The corresponding LEED pattern is reproduced in figure 4(c), where the hexagonal symmetry is clearly recognized. The lattice constant of the HEX-II oxide phase as derived from STM and LEED is $b_1 = b_2 = 6.0$ Å ± 0.05 Å, and the LEED simulation in figure 4(d) confirms this lattice model.

*3.2 The MnO$_x$ phases at "intermediate" oxygen pressures: $5 \times 10^{-7} > p_{O2} > 1 \times 10^{-7}$ mbar*

At "intermediate" oxygen pressures four different MnO$_x$ phases have been observed. At the higher pressure end of this regime, the so-called *stripe* phase is found at the surface (figures 5(a) and (b)), which is often coexistent with the c(4×2) structure (figures 5(c) and (d)). The *stripe* phase occurs in two different appearances, which are caused by a different arrangement of anti-phase domain boundaries. In figure 5(a) bands of parallel stripes are formed by the domain boundaries running at an angle of 60° with respect to the stripes (see arrows on the figure); this results in *w*-shaped structure motifs. In figure 5(b) the anti-phase domain lines run in orthogonal directions, resulting in small square or rectangular domains. The separation between the stripes is ~5.5 Å, which is 2 × $a_{Pd}$, while along the stripes maxima separated by ~5.4 Å have been recognized in some STM images. The *stripe* phase thus appears to be a quasi-hexagonal phase; this is supported by the analysis of the LEED pattern, which is discussed below.

The c(4×2) structure is the central oxide phase in this pressure regime. It is a commensurate structure, characterized by $\mathbf{b_{1,2}}$ = 6.15 Å or by the transformation matrix $\mathbf{M}$ = [2, -1 / 2,1]. Under carefully controlled kinetic conditions the c(4×2) structure forms a single phase wetting layer as shown in the STM images of figures 5(c) and (d). In figure 5(c), well-ordered c(4×2) domains, in two orthogonal directions, are separated by disordered phase boundaries. The latter presumably act as a relief of interfacial strain. The c(4×2) $MnO_x$ structure observed here is similar to the c(4×2) Ni oxide monolayer structure on Pd(100) as reported by Agnoli et al. [17]. The latter has been interpreted in terms of a compressed NiO-type monolayer with a c(4×2) Ni vacancy array, yielding an overall stoichiometry of $Ni_3O_4$. We suggest that a similar model may also be applicable in the case of c(4×2) $MnO_x$ [12].

The LEED pattern displayed in figure 6(a) is from a surface, which contained mainly the *stripe* oxide phase; due to limited long-range order the pattern is diffuse, but the relevant features can still be recognized. We argue that the *stripe* structure can be created by a distortion of the c(4×2) lattice. Figure 6(b) illustrates this point: by a uniaxial compression of the c(4×2) unit cell of 5% along the $\mathbf{a_2}$ direction (arrows in figure 6(b)) a transformation matrix of $\mathbf{M}$ = [2, -0.95 / 2, +0.95] is obtained. The structure according to this transformation matrix yields the diffraction pattern of figure 6(c) [16], which reproduces remarkably well the experimental LEED pattern of figure 6(a).

At the lower pressure end of the "intermediate" pressure regime the c(4×2) phase is often observed in coexistence with the so-called *chevron* structures (figure 7). The name for these structures has been deduced from the "chevron-like" structure motifs as seen in figure 7(a), which are derived from different domains separated by anti-phase domain boundaries. There are two structures, *chevron I* and *chevron II*, corresponding to the transformation matrices $\mathbf{M}$ = [2, -1 / 3, 2] and $\mathbf{M}$ = [2, -1 / 5, 3], respectively. In STM, these structures consist of rows of maxima (figure 7(b)), with unit cell vectors $\mathbf{b_1}$ = 6.15 Å, specifying the common distance of maxima along the rows, and $\mathbf{b_2}$ = 9.92 Å and 16.04 Å for *chevron I* and *II*, respectively; the latter define the distances between the rows. The unit cells for the two structures *I* and *II* are indicated on figure 7(b). The phase boundaries between the two *chevron* structures and the c(4×2) structure are smooth and continuous (see dashed line in figure 7(c)), suggesting that these phases are structurally closely related. Indeed, the unit cells of the *chevron* structures can be generated from the c(4×2) structure by a simple extension of the **b$_2$** unit cell vectors to adjacent anti-phase positions as illustrated in figure 8(b). Figure 8(a) displays the experimentally observed LEED patterns of the c(4×2), *chevron I* and *II* structures, whereas figure 8(c) shows the computer simulated patterns derived from the structure models of figure 8(b) [16]. There is excellent agreement between the experimental and simulated patterns. The formation of the *chevrons* from the c(4×2) structure can be rationalized in terms of a Mn vacancy propagation model as discussed in a forthcoming publication [12].

*3.3 The MnO$_x$ phases in the "oxygen-poor" regime: $1 \times 10^{-7} > p_{O2} > 5 \times 10^{-8}$ mbar*

There is a smooth transition from the c(4×2) and *chevron* structures of the intermediate oxygen pressure regime to the structures in the "oxygen-poor" regime: *labyrinth* and *waves*, which are named according to their STM pattern. In the STM image of figure 9(a) three large terraces are seen, where the upper left and the lower right terraces are covered by the *labyrinth* (L) structure, whereas the middle terrace is coated by the *waves* (W). The lines that run parallel to the step edges belong to the c(4×2) structure. Figure 9(c) shows a close-up of the *labyrinth* structure; at the step edge the c(4×2) structure is clearly recognized here. The *labyrinth* structure is described by the lattice vectors **b$_1$** = 13.5 ± 0.5 Å and **b$_2$** = 11.0 ± 0.5 Å, θ = 85 ± 2°, the unit cell is rotated by α = 21 ± 2° with respect to the Pd [011] direction. The *waves* structure is illustrated in the STM images of figure 9(b) and in the LEED pattern of figure 9(d). The unit cell is characterized by **b$_1$** = 13.8 ± 0.2 Å, **b$_2$** = 34 ± 1 Å, θ = 90 ± 2°, which corresponds approximately to a (5×12) superstructure with respect to the Pd(100) substrate. This is compatible with the LEED pattern of figure 9(d). From the STM images, the periodicity along the **b$_2$** direction may also be twice as large, supporting a (5×24) superstructure.

In figure 9(b) an embedded island is visible in the centre, where a different structure can be perceived. This structure with a hexagonal symmetry is formed best at the low oxygen pressure end of this regime or by the oxidation of Mn atoms that have segregated from the Pd substrate bulk to the surface (after prolonged use of the Pd crystal as a substrate for Mn oxide growth, some Mn atoms become dissolved in the Pd bulk). This HEX-III structure is shown more clearly in the STM image of figure 10(a), with the corresponding LEED pattern in figure 10(b). The HEX-III structure is described by the transformation matrix **M** = [0, 2 / √3, 1] with respect to Pd(100), but corresponds also to a (√3 × √3)R30° superstructure with respect to a MnO(111) surface. It is possible that this structure is an interfacial phase that mediates the growth of MnO(111) oriented films, which has been reported previously [11].

*3.4 Mn 2p and O 1s XPS core level spectra*

In figure 11(a) we show two representative Mn $2p_{3/2}$ XPS core level spectra of the HEX-I and the *waves* phases. The Mn 2p core level photoemission lines of Mn oxides are broad and display a complex shape due to multiplet splitting, correlation and configuration interaction effects in the final state [18]. Moreover, in oxide nanolayers on metals the screening from the substrate influences the determination of the cation oxidation state by comparison with bulk oxide XPS data. Thus, the absolute Mn $2p_{3/2}$ binding energies measured in figure 11(a) at around 640 – 641.5 eV, i.e. at values typical for MnO, are perhaps not too meaningful. However, the shift of the Mn 2p binding energy of 0.8 eV between the HEX-I and the *waves* structure is significant and justifies the separation of the $MnO_x$ phase diagram into an "oxygen-rich" region with phases of higher oxidation state and into an "oxygen-poor" region with lower oxidation state Mn oxides. Direct confirmation of this conclusion is provided by the O 1s spectra reported in figure 11(b). Here, the O 1s core level intensity is strongly reduced when moving from the HEX-I to the *waves* phase. Although photoelectron diffraction may contribute to this effect, its contribution is typically much smaller than the intensity reduction observed in figure 11(b). The experimental evidence therefore suggests a markedly different stoichiometry for the "oxygen-rich" and "oxygen-poor" regions. Interestingly, the broad O 1s line-shape for the HEX-I phase points to the presence of two oxygen components, one centred at 529.1 eV and the other chemically shifted to higher binding energy by roughly 0.4 eV. This is indicative of an oxide layer with differently coordinated O ions, as confirmed by the structural model that will be presented in reference [12].

4. Conclusions

The formation of two-dimensional $MnO_x$ nanolayer phases on Pd(100) has been studied by STM, LEED and XPS as a function of the oxygen pressure during preparation at a substrate temperature of ~700 K. The complex phase diagram of 2D Mn oxide structures on Pd(100) as displayed in figure 12 in form of the stability region of oxide phases versus the chemical potential of oxygen $\mu_O$ summarizes the results of this investigation. Nine major 2D oxide nanophases have been detected in the accessible region of the parameter space. In the so-called "oxygen-rich" region, at $\mu_O$ = -1.15 – 1.35 eV, the hexagonal phases HEX-I and HEX-II have been observed, the former in two modifications as a *distorted* and *undistorted* hexagonal structure. At "intermediate" chemical potentials, $\mu_O$= -1.35 – 1.45 eV corresponding roughly to the $10^{-7}$ mbar oxygen pressure region, the c(4×2) $MnO_x$ structure is the central phase. The other structures found in this region are the *stripe*, the *chevron I* and the *chevron II* structures. The latter overlayers are closely related to the c(4×2) phase as indicated by their frequent coexistence at the surface and the smooth boundary regions between them. At "low" chemical potentials of oxygen, $\mu_O < -1.45$ eV, the *waves* and *labyrinth* structures are observed, which have complex unit cells and for which the connection to the phases at intermediate oxygen potentials is less clear. At the lowest $\mu_O$, an additional hexagonal phase HEX-III is formed with unit cell dimensions that are related to those of the MnO(111) face. Finally, we mention that the $MnO_x$ phases described above are the ones which have been reproducibly observed many times. There are several intermediate phases which have been seen only occasionally and which seem to be metastable and only stabilized under particular kinetic conditions.

**Acknowledgements**

This work has been supported by the Austrian Science Funds and the EU STREP programme GSOMEN. Technical assistance of the staff at the Beamline I311 (MAX II, Lund) is gratefully acknowledged.

**Figure captions:**

Figure 1: HEX-I distorted MnOx phase: (a) Large-scale STM image (2000Å×2000Å, U= -1 V, I=0.15nA). (b) LEED pattern (E=96 eV). (c) High-resolution STM images (100Å×100Å, U=+0.5V, I=0.13nA). Inset (20Å×17Å, U=+0.6V, I=0.15nA). (d) Line profile, taken along the line indicated in panel (c).

Figure 2: HEX-I distorted MnOx phase: (a) Real lattice model: the square mesh corresponds to the Pd lattice with unit cell vectors ($a_1, a_2$). The black dots represent the HEX-I distorted MnOx lattice with unit cell vectors ($b_1, b_2$). (b) The corresponding reciprocal lattice pattern [16]. (c) A hard sphere model of the HEX-I distorted lattice superimposed on the Pd(100) lattice, reproducing the Moiré-pattern contrast and periodicity of the STM image in figure 1(c).

Figure 3: HEX-I undistorted MnOx phase: (a) LEED pattern (E=90 eV). (b) Real lattice model: the square mesh corresponds to the Pd lattice with unit cell vectors ($a_1, a_2$). The black dots represent the HEX-I undistorted MnOx lattice with unit cell vectors ($b_1, b_2$). (c) The corresponding reciprocal lattice pattern [16].

Figure 4: HEX-II MnOx structure: (a) STM image (300Å×300Å, U=+1.5V, I=0.1nA). (b) STM image (150Å×150Å, U=+1.5V, I=0.1nA). (c) LEED pattern (E=80 eV). (d) Simulated LEED (reciprocal space) pattern [16].

Figure 5: STM images of the *stripe* MnOx phase with two different types of anti-phase domain boundaries: (a) linear (300Å×170Å, U=+1.4V, I=0.34nA) and (b) square-like (300Å×300Å, U=+0.6V, I=0.2nA). The arrows in figure 5(a) mark the position of the anti-phase domain boundaries. STM images of the c(4×2) structure: (c) (500Å×500Å, U=+0.5V, I=0.2nA) and (d) (200Å×200Å, U=+0.5V, I=0.2nA).

Figure 6: *Stripe* MnOx phase: (a) LEED pattern (E=60 eV). (b) Real lattice model: the square mesh corresponds to the Pd lattice with unit cell vectors (**a$_1$, a$_2$**). The black dots represent the *stripe* MnOx lattice with unit cell vectors (**b$_1$, b$_2$**). (c) The corresponding reciprocal lattice pattern [16].

Figure 7: STM images of the *chevron* MnOx phases: (a) 300Å×300Å, U=+1.0V, I=0.2nA. (b) 150Å×150Å, U=+1.2V, I=0.26nA). The unit cells of the *chevron I* and *chevron II* structures are indicated. (c) STM image (200Å×200Å, U=+1.0V, I=0.1nA) showing an area of coexisting c(4×2) and *chevron* structures, separated by a smooth phase boundary (white dashed line).

Figure 8: LEED patterns (a), real lattice models (b) and reciprocal lattice patterns (c) of the c(4×2) (left panel), *chevron I* (middle panel) and *chevron II* (right panel) structures. The respective LEED energies in (a) are 116 eV, 108 eV and 60 eV.

Figure 9: (a) STM image (500Å×500Å, U=+1.5V, I=0.1nA) of *labyrinth (L)* and *wave (W)* MnOx structures covering neighbouring terraces on the Pd(100) surface. (b) High-resolution STM images of the *wave* structure (300Å×300Å, U=+0.6V, I=0.3nA). Inset (200Å×200Å, U=+1.4V, I=0.3nA). (c) High-resolution STM images of the *labyrinth* structure (300Å×300Å, U=+0.6V, I=0.3nA). The terrace areas close to the step edges are covered by the c(4×2) phase. Inset (90Å×90Å, U=+0.1V, I=0.45nA). (d) LEED pattern of the *wave* structure (100 eV).

Figure 10: (a) STM image (90Å×90Å, U=+3.3 mV, I=1.0 nA) and (b) LEED pattern (85 eV) of the *HEX-III* MnOx phase.

Figure 11: (a) Mn 2p$_{3/2}$ and (b) O 1s core level spectra of the *HEX-I* (top curves) and *waves* (bottom curves) MnOx phases.

Figure 12: Schematic phase diagram of the 2D Mn oxides, presented as a function of the oxygen pressure p(O$_2$) and of the oxygen chemical potential μ$_O$. The nominal coverage of Mn on Pd(100) is 0.75 ML.

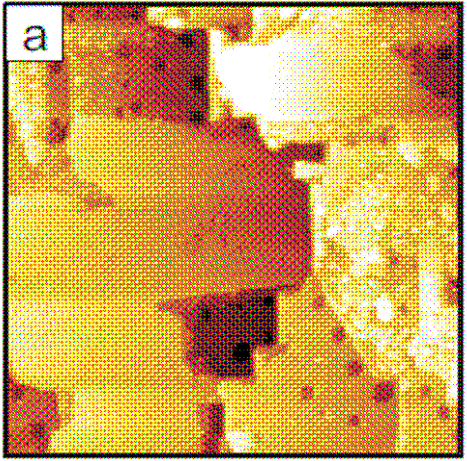 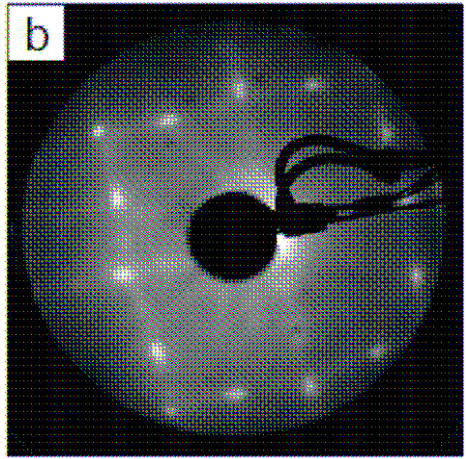
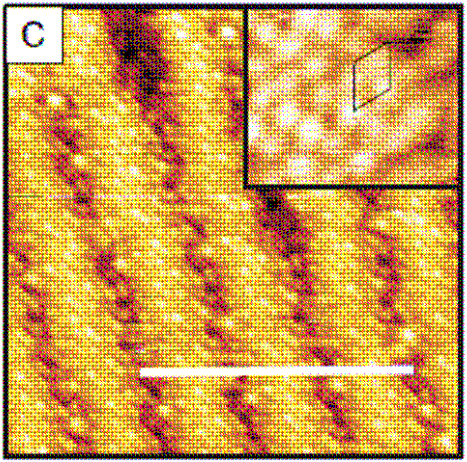 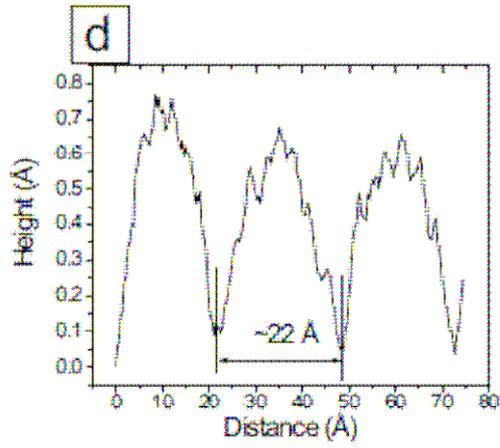

Figure 1

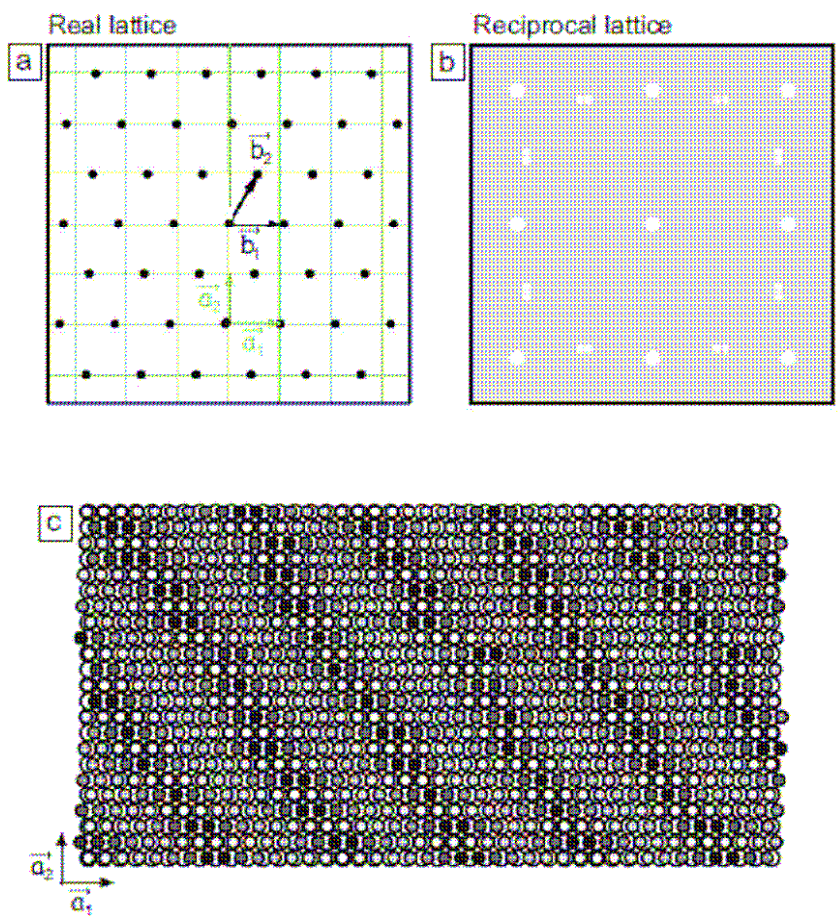

Figure 2

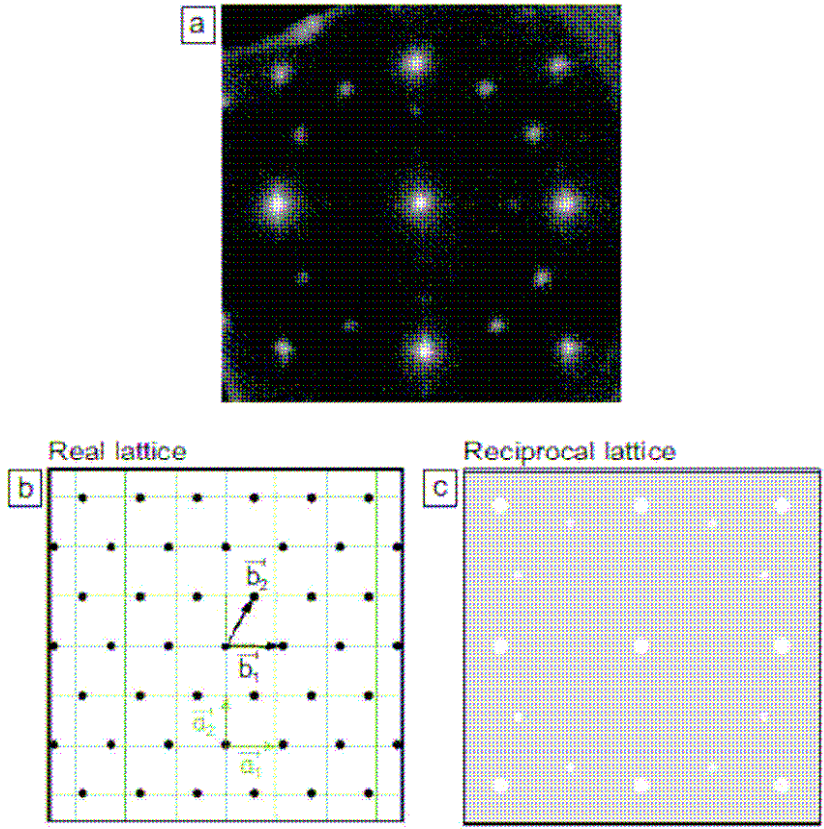

Figure 3

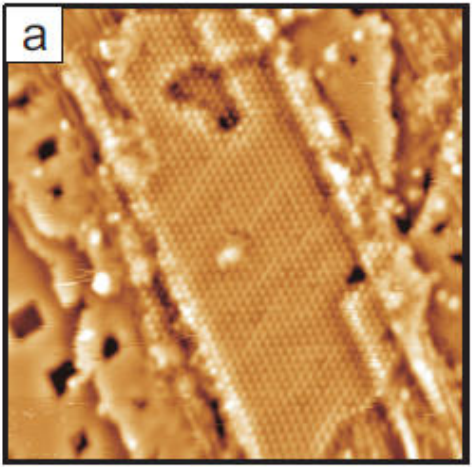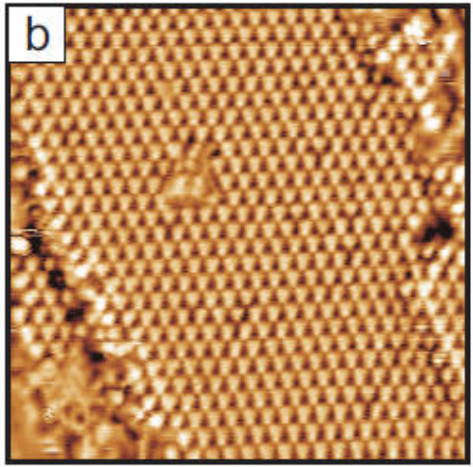
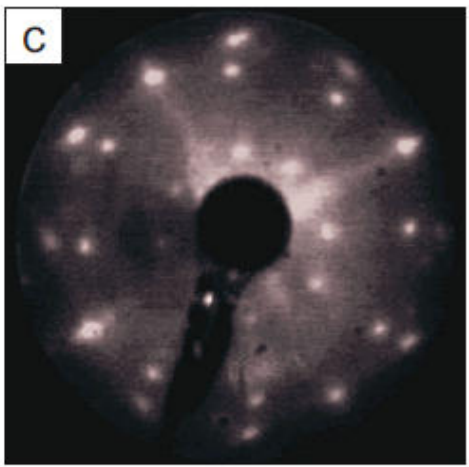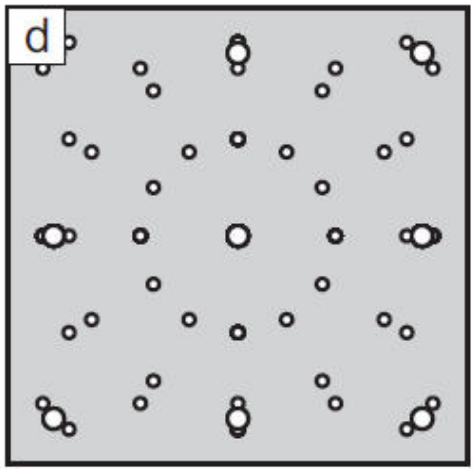

Figure 4

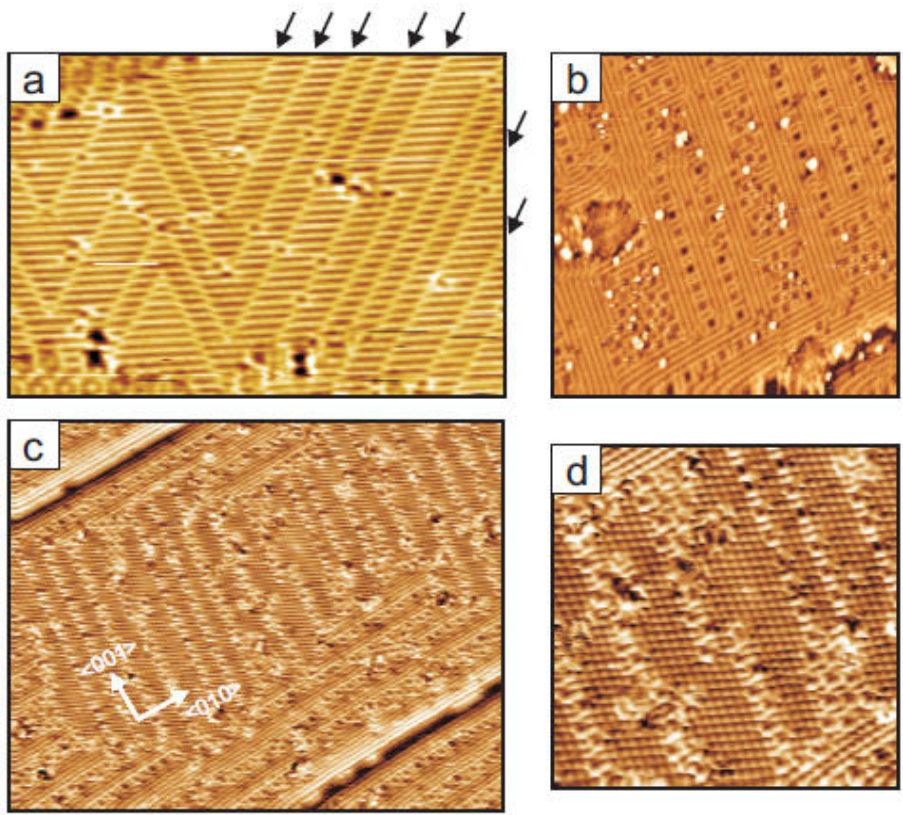

Figure 5

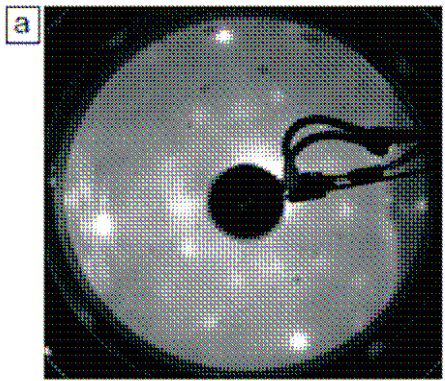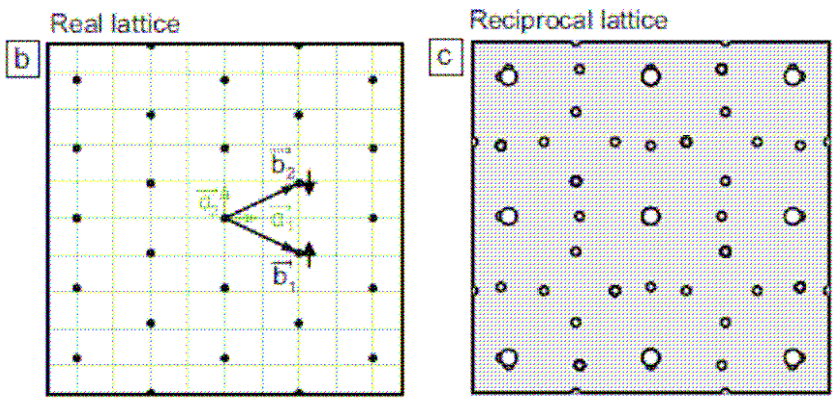

Figure 6

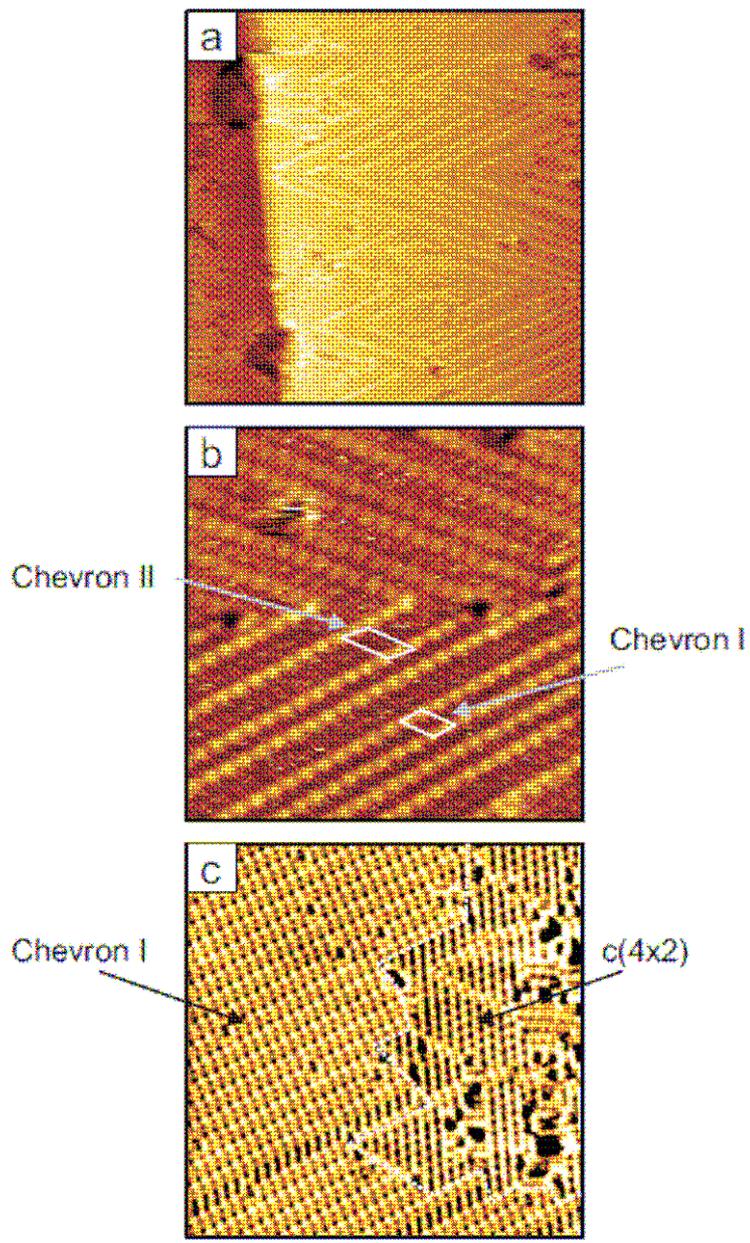

Figure 7

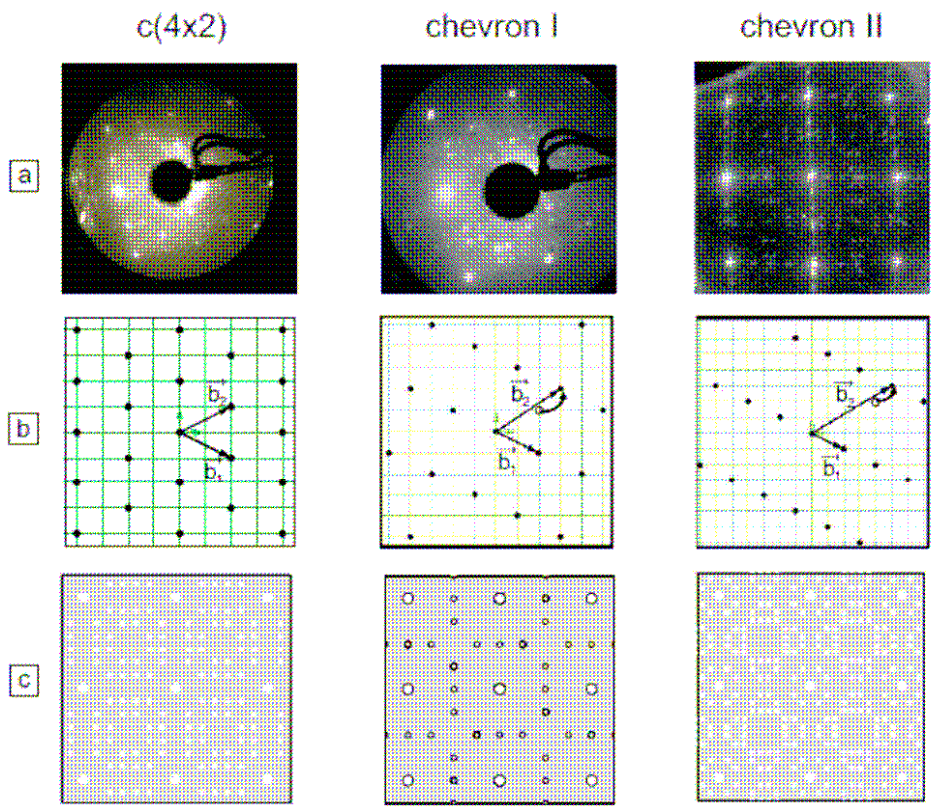

Figure 8

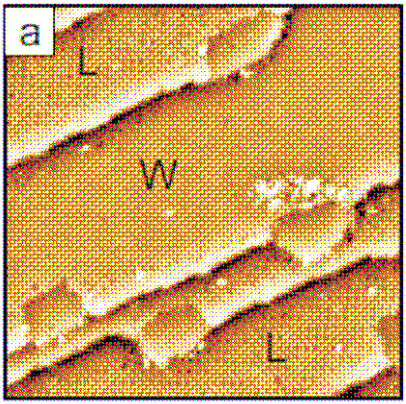 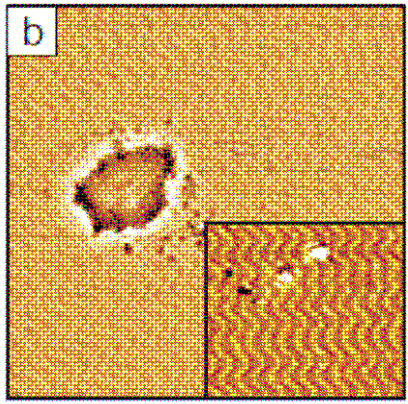
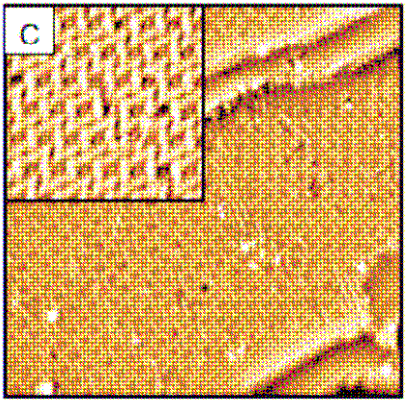 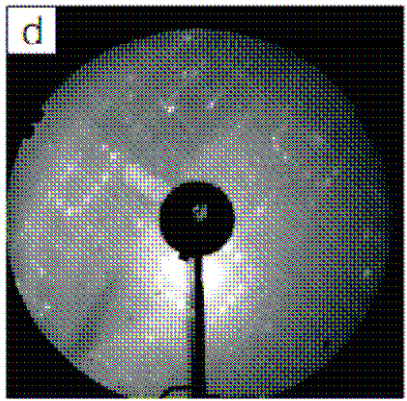

Figure 9

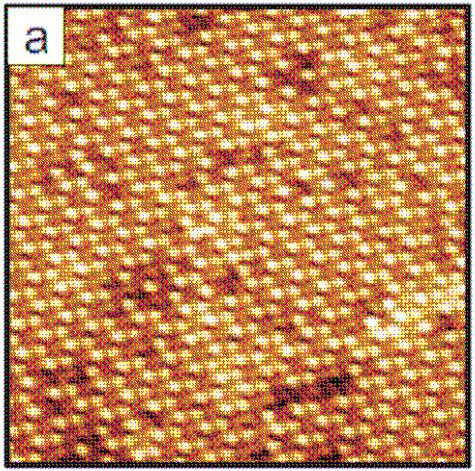
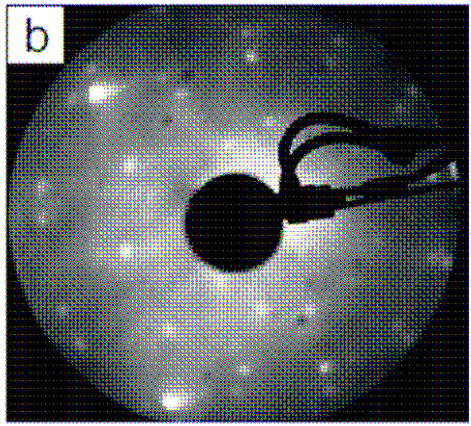

Figure 10

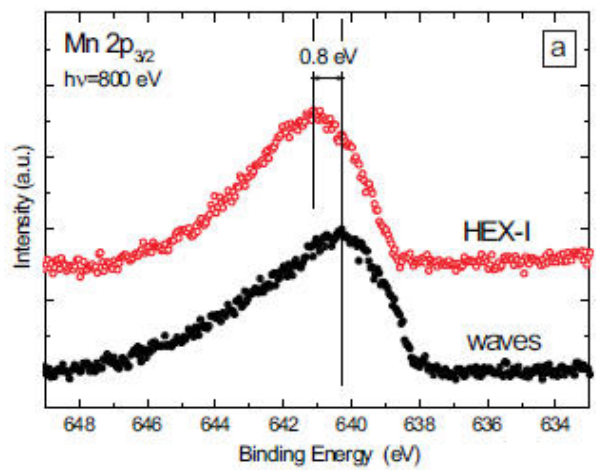
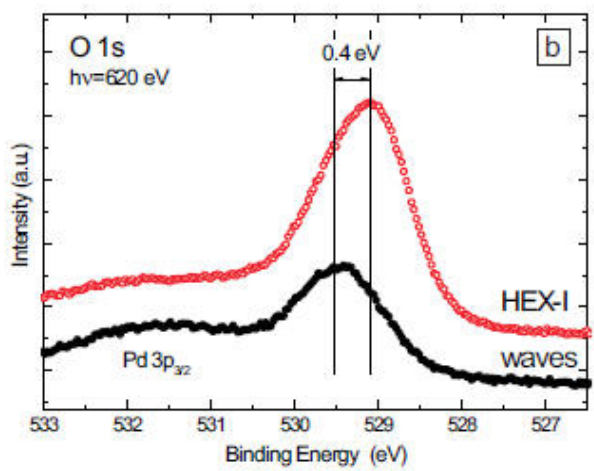

Figure 11

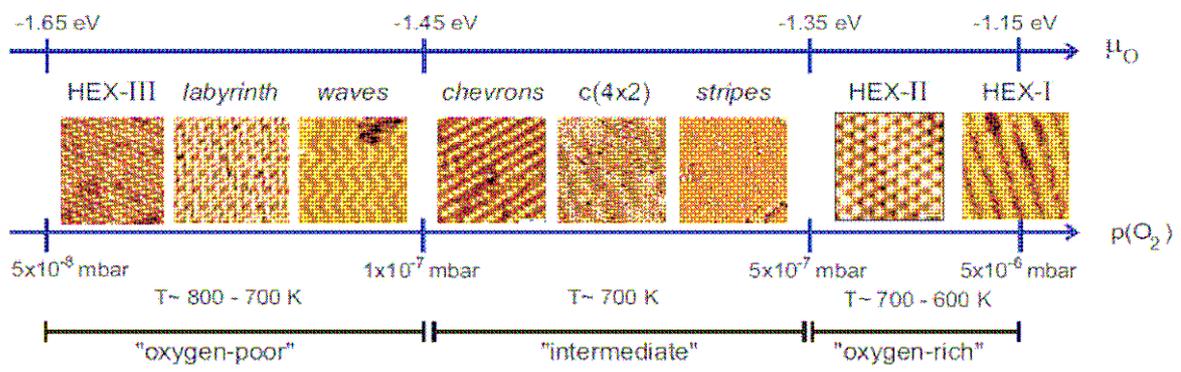

Figure 12